\documentclass[preprint,showpacs,aps,prd]{revtex4-1}
\usepackage{epsfig}
\usepackage{psfrag}
\usepackage{graphicx}
\usepackage{titlesec}
\usepackage{hyperref}
\usepackage{graphics}
\usepackage{amsmath}
\usepackage{color}
\usepackage{amssymb}
\usepackage[title]{appendix}

\begin{document}


\title{Considerations on electromagnetic radiation in Podolsky electrodynamics within a Lorentz-symmetry violating framework}

\author{J. P. Ferreira} \email{ferreira\_jp@unifei.edu.br}
\affiliation{Instituto de F\'{\i}sica e Qu\'{\i}mica, Universidade Federal de Itajub\'a, MG, Brasil}

\author{Patricio Gaete} \email{patricio.gaete@usm.cl} 
\affiliation{Departamento de F\'{i}sica and Centro Cient\'{i}fico-Tecnol\'ogico de Valpara\'{i}so-CCTVal,
Universidad T\'{e}cnica Federico Santa Mar\'{i}a, Valpara\'{i}so, Chile}

\author{Jos\'e Abdalla Helay\"{e}l-Neto}\email{helayel@cbpf.br}
\affiliation{Centro Brasileiro de Pesquisas F\'{i}sicas (CBPF), Rio de Janeiro, RJ, Brasil} 

\date{\today}

\begin{abstract}
We explore how higher-order derivative terms impact a physical observable within a Lorentz-Symmetry Violating (LSV) framework. We specifically examine Podolsky electrodynamics coupled with the Carroll-Field-Jackiw model (CFJ). Our analysis shows how vacuum and wave propagation are affected by the interaction of LSV parameters with properties of Podolsky electrodynamics. In this scenario, our results indicate that the new electromagnetic vacuum is birefringent, dichroic, and dispersive. We also point out that the higher-derivative term in the presence of the CFJ space-time anisotropy induces an oscillation mechanism
between the photon, which acquires a non-vanishing mass, and the
massive vector mode carried by the higher derivative.
We then investigate the electromagnetic radiation emitted by a moving charged particle interacting with this new medium. Our analysis reveals that the refractive vacuum through which the charged particle travels influences the electromagnetic radiation, leading to a profile similar to the Cherenkov effect.
\end{abstract}


\maketitle

\section{Introduction}
Theories involving higher-order derivatives, such as Podolsky's electrodynamics \cite{Bopp, Podolsky, Schwed, Gaete, Galvao, Accioly}, have attracted significant attention due to their potential to achieve finite theories at short distances. In this $U(1)$ gauge theory, a quadratic term in the divergence of the field strength tensor is added to the free Lagrangian of the $U(1)$ sector. This leads to exciting features, such as a finite electron self-energy and a regular point charge electric field at the origin. These theories have also attracted interest in connection with supersymmetric and string theories \cite{West, Polyakov}. In addition to Podolsky's electrodynamics, another exciting theory is known as Lee and Wick's electrodynamics \cite{Lee, Wick}. This theory introduces a dimensional-6 operator containing higher derivatives to the free Lagrangian of the $U(1)$ sector in the $U(1)$ gauge theory. Following this proposal, considerable recent attention has been paid to studying modifications to the Standard Model, which stabilizes the Higgs mass against quadratically divergent corrections leading to the Lee-Wick Standard Model \cite{Grinstein}. Along the same line, we also recall that a Podolsky term also appears in the context of non-commutative electrodynamics \cite{GaeteSpa} and in the study of an effective Lagrangian for QED \cite{Donoghue}.

In this perspective, we also point out that vacuum polarization in quantum electrodynamics (QED) arises from the polarization of virtual electron-positron pairs, leading to nonlinear interactions between electromagnetic fields \cite{Euler, Adler, Costantini, Biswas, Tommasini, Ferrando}. It is worth mentioning that while
 there has been considerable progress in this area, confirming 
the phenomenon of vacuum polarization remains elusive. Euler and Heisenberg's work \cite{Euler} provides a notable example, where they computed an effective nonlinear electromagnetic theory stemming from the interaction of photons with virtual electron-positron pairs in a vacuum. In this connection, the study of quantum vacuum nonlinearities and their effects, such as vacuum birefringence and vacuum dichroism, have been of great interest since their inception. These fascinating quantum properties of light have driven increased interest in experimental research \cite{Bamber, Burke, Pike}. For instance, both the PVLAS Collaboration \cite{Ejlli} and the more recent  ATLAS Collaboration have reported direct detection of the light-by-light scattering in LHC Pb–Pb collisions \cite{Aaboud, Enterria}. Moreover, advancing laser facilities have led to various proposals investigating quantum vacuum nonlinearities.

On the other hand, let us also recall that the Standard Model of particle physics relies on Lorentz invariance, which is an exact symmetry and a key characteristic of modern physics. \cite{Kostelecky1, Kostelecky2} Any potential violation of Lorentz symmetry could indicate the presence of new physics. Consequently, extensive research has been conducted, particularly in string theories, where the violation of Lorentz and CPT symmetries at a fundamental scale could occur spontaneously, addressing the theoretical challenges within the framework of quantum gravity.
Specific theories, such as loop quantum gravity and string theories \cite{Samuel1, Samuel2, Gambini, Alfaro}, which aim to offer a quantum-consistent description of gravity, anticipate the breakdown of Lorentz and CPT around the Planck scale. Violating these symmetries at lower energies might manifest as modifications in the dispersion relations of photons.

 In this way, the physical consequences at low energies of Lorentz and CPT symmetry violation can be studied in the context of effective field theories, which can provide valuable information to constrain a more fundamental theory. Thus, the Standard Model Extension (SME) provides a suitable framework for testing the low-energy manifestations of LSV and allows for the realization of spontaneous LSV.

Based on these observations, we aim to explore another aspect of Podolsky's electrodynamics within a Lorentz-symmetry-violating framework. Specifically, we want to assess how higher-order derivative terms coupled with the Carroll-Field-Jackiw model (CFJ) affect a physical observable.
Therefore, we aim to understand better the potential observational signatures introduced by this new electromagnetic vacuum, focusing on investigating
birefringence, dichroism, and electromagnetic radiation.

To conduct these studies, we start by calculating the photonic dispersion relations to understand how the higher derivative term coupled with the Carroll-Field-Jackiw model (CFJ) affects the constitutive properties of the vacuum. In addition, we determine the refractive indices, showing that the new medium is birefringent, dichroic, and dispersive. Based on these findings, we examine the electromagnetic radiation a moving charge produces when interacting with this new dispersive medium. We will calculate the radiated energy using the standard Poynting vector approach. Our analysis reveals the importance of the higher derivative terms in triggering the radiated energy.

Our work is organized according to the following outline: Section 2 introduces the model we are considering and provides the dispersion relations and the refractive indices. Section 3 discusses electromagnetic radiation and shows that the radiation obtained is similar to that of the Cherenkov effect. Finally, in Section 4, we present some closing remarks.

In our conventions, the signature of the metric is ($+1,-1,-1,-1$).

\section{Model under consideration: dispersion relations and refractive indices}


We start by providing a summary of the model under consideration. The following Lagrangian density describes the model under consideration:
\begin{equation}
{\cal L} =  - \frac{1}{4}{F_{\mu \nu }}{F^{\mu \nu }} +\frac{l^2}{2} (\partial_{\alpha}F^{\beta\alpha})^2 + \frac{1}{4}{\varepsilon ^{\mu \nu \kappa \lambda }}{v_\mu }{A_\nu }{F_{\kappa \lambda }}, \label{PodCFJ01}
\end{equation}
which can be rewritten as
\begin{equation}
{\cal L} =  - \frac{1}{4}{F_{\mu \nu }}{F^{\mu \nu }} - \frac{l^{2}}{{4}}{F_{\mu \nu }}\,\Box \,{F^{\mu \nu }} + \frac{1}{4}{\varepsilon ^{\mu \nu \kappa \lambda }}{v_\mu }{A_\nu }{F_{\kappa \lambda }}, \label{PodCFJ05}
\end{equation} 
here $\Box  \equiv {\partial _\mu }{\partial ^\mu }$ and ${v_\mu}$ is an arbitrary four-vector that selects a preferred direction in space-time, with mass dimensions in natural units. Furthermore, $l$ is a real constant with a dimension of length bounded by
$l < 4.7 \times 10^{-18}$ m \cite{Tony}.

The field equations are given by
\begin{eqnarray}
\left( {1 + {l^2}\,\Box } \right)\nabla  \cdot {\bf E} - c\, {\bf v} \cdot {\bf B}= 0, \nonumber\\
\nabla  \cdot {\bf B} = 0, \nonumber\\
\nabla  \times {\bf E} + \frac{{\partial {\bf B}}}{{\partial t}} = 0, \nonumber\\
\left( {1 + {l^2}\,\Box } \right)\left[ {\nabla  \times {\bf B} - \frac{{1}}{c^{2}}\frac{{\partial {\bf E}}}{{\partial t}}} \right] - {v^0}\, {\bf B} + \frac{1}{c} {\bf v}  \times {\bf E}= 0.  
\label{PodCFJ10}
\end{eqnarray}

We are now focusing on another aspect of vacuum electromagnetic properties. To be more precise, we aim to understand the connection between the parameters of high-order derivative terms sector and LSV parameters, such as $v^{\mu}$ for CFJ. In the following discussion, we will examine this aspect more explicitly.

The first step in this direction is to analyze the dispersion relations (DRs) for an electromagnetic wave traveling through this new electromagnetic vacuum. 
To achieve this, we will decompose the fields $\bf E$ and $\bf B$ into plane waves, namely, 
\begin{equation}
{\bf E} = {{\bf E}_0}{e^{i\left( {k \cdot x - \omega t} \right)}} + c.c.,  \label{PodCFJ15}
\end{equation}
and
\begin{equation}
{\bf B} = {{\bf B}_0}{e^{i\left( {k \cdot x - \omega t} \right)}} + c.c.
.  \label{PodCFJ20}
\end{equation}

Making use of the previous field equations, we find that 
\begin{equation}
{M_{ij}}{E_{0j}} = 0,  \label{PodCFJ25}
\end{equation}
where the matrix $M_{ij}$ is given by
\begin{equation}
{M_{ij}} = \left( {1 - {l^2}{k^2}} \right)\left( {\frac{{{\omega ^2}}}{{{c^2}}} - {{\bf k}^2}} \right){\delta _{ij}} + \left( {1 - {l^2}{k^2}} \right){k_i}{k_j} - i{\varepsilon _{ijk}}{\left( {{v^0}{\bf k} - \frac{\omega }{c}{\bf v}} \right)_k}. \label{PodCFJ30}
\end{equation}
It may be noted here that the matrix ${M_{ij}}$ is of the type ${M_{ij}} = \alpha {\delta _{ij}} + \beta {u_i}{w_j} + \gamma {\varepsilon _{ijk}}{\zeta _k}$, whose determinant is given by $\det M = {\alpha ^3} + {\alpha ^2}\beta \left( {{\bf u} \cdot {\bf w}} \right) + \alpha {\gamma ^2}{{\pmb \zeta}^2} + \alpha \beta w\left( {{\bf u} \times {\bf w}} \right) \cdot {\pmb \zeta}  + \beta {\gamma ^2}\left( {{\bf u} \cdot { \pmb \zeta} } \right)\left( {{\bf w} \cdot {\pmb \zeta} } \right)$.
Making use of the condition $\det M=0$, we can determine that the corresponding dispersion relations are given by
\begin{eqnarray}
{\left( {1 - {l^2}{k^2}} \right)^2}{\left( {\frac{{{\omega ^2}}}{{{c^2}}} - {{\bf k}^2}} \right)^3} + {\left( {1 - {l^2}{k^2}} \right)^2}{\left( {\frac{{{\omega ^2}}}{{{c^2}}} - {{\bf k}^2}} \right)^2}{{\bf k}^2} - \left( {\frac{{{\omega ^2}}}{{{c^2}}} - {{\bf k}^2}} \right){\left( {{v^0}{\bf k} - \frac{\omega }{c}{\bf v}} \right)^2} \nonumber\\
- {\left( {{v^0}{{\bf k}^2} - \frac{\omega }{c}{\bf v} \cdot {\bf k}} \right)^2} = 0.  \label{PodCFJ35}
\end{eqnarray}

After some manipulations, the previous equation can be brought to the covariant form
\begin{equation}
{\left( {1 - {l^2}{k^2}} \right)^2}{k^4} + {v^2}{k^2} - {\left( {v \cdot k} \right)^2} = 0.\label{PodCFJ40}
\end{equation}

It is easy to verify the internal consistency of our procedure. By excluding the Podolsky term ($l=0$), equation (\ref{PodCFJ40}) simplifies to one found in the CFJ model, 
\begin{equation}
{k^4} + {v^2}{k^2} - {\left( {v \cdot k} \right)^2} = 0.  \label{PodCFJ45}
\end{equation}

Before going on, it is appropriate to observe here that by considering the special case  $v^{0}=0$, one can search for the effective mass $m_\gamma$ acquired by the photon. In fact, from the previous dispersion relation with $v^{0}=0$ and ${\bf k}=0$, we obtain
\begin{equation}
{l^4}{\left( {\frac{{{\omega ^2}}}{{{c^2}}}} \right)^3} - 2{l^2}{\left( {\frac{{{\omega ^2}}}{{{c^2}}}} \right)^2} + \frac{{{\omega ^2}}}{{{c^2}}} - {{\bf v}^2} = 0. \label{PodCFJ50}
\end{equation}
With this in hand, and using $\hbar \omega  = E_\gamma ^{rest} = {m_\gamma }{c^2}$, to order ${\cal O}(l^{2})$ the masses photon are found to be ${m_\gamma } = \frac{{\hbar |{\bf v}|}}{c}$ and ${m_\gamma } = \frac{\hbar }{{c\sqrt 2 }}\frac{1}{l}\sqrt {1 - {l^2}{{\bf v}^2}}$. Evidently, the first mass corresponds to the Maxwell case, while the second one exhibits the contribution of the Podolsky term.

We now proceed to calculate the refractive indices. These can also be determined using the previous dispersion relation. We will start by noting that if $v=0$, equation (\ref{PodCFJ40}) reduces to 
\begin{equation}
{k^2}\left( {1 - {l^2}{k^2}} \right) = 0. \label{PodCFJ55}
 \end{equation}
An immediate consequence of this result is providing two expressions for the refractive indices. The first corresponds to the usual Maxwell case $(k^2=0)$ leading to the refractive index $n(\omega)=1$. Whereas, from $\left( {1 - {l^2}{k^2}} \right) = 0$, we obtain
 \begin{equation}
 {n^2}\left( \omega  \right) \equiv \frac{{{c^2}|{\bf k}{|^2}}}{{{\omega ^2}}} = 1 - \frac{{{c^2}}}{{{\omega ^2}{l^2}}}. \label{PodCFJ60}
 \end{equation}
Consequently, the new electromagnetic vacuum behaves as a birefringent, dichroic, and dispersive medium.
 
Within this context, we also point out here that by using ${\bf D} = \frac{{\partial {\cal L}}}{{\partial {\bf E}}}$, ${\bf H} =  - \frac{{\partial {\cal L}}}{{\partial {\bf B}}}$, ${\bf D} = \varepsilon \,{\bf E}$ and ${\bf H} = {\mu ^{ - 1}}{\bf B}$, we obtain the following expressions for the vacuum permittivity and vacuum permeability:
 \begin{equation}
 \varepsilon  = 1 + {l^2}\,\Box,\label{PodCFJ61}
 \end{equation}
 and
 \begin{equation}
\mu  = \frac{1}{{\left( {1 - {l^2}\,\Box } \right)}}. \label{PodCFJ62}
 \end{equation}
Since $l$ is very small ($\sim10^{-18}$ m), it is easy to see that up to order
$l^{2}$ (${\cal O}\left( {{l^2}} \right)$) we have $\varepsilon  \equiv \mu$.
In the following Section, we will calculate the electromagnetic radiation to ${\cal O}\left( {{l^2}} \right)$.
 
Next, we also notice that for the case ${\bf v} \ne 0$ e $v^{0}=0$, the dispersion relation becomes
\begin{equation}
\left[ {1 - 2{l^2}\left( {\frac{{{\omega ^2}}}{{{c^2}}} - {{\bf k}^2}} \right) + {l^4}{{\left( {\frac{{{\omega ^2}}}{{{c^2}}} - {{\bf k}^2}} \right)}^2}} \right]{\left( {\frac{{{\omega ^2}}}{{{c^2}}} - {{\bf k}^2}} \right)^2} - {{\bf v}^2}\left( {\frac{{{\omega ^2}}}{{{c^2}}} - {{\bf k}^2}} \right) - {\left( {{\bf v} \cdot {\bf k}} \right)^2} = 0. \label{PodCFJ65}
\end{equation} 
 
Upon further simplification, at order ${\cal O}(l^{2})$, we find the following refractive indices
\begin{eqnarray}
{n^2}\left( \omega  \right) \equiv {c^2}\frac{{|{{\bf k}^2}|}}{{{\omega ^2}}} &=& 1 - \frac{{{c^2}}}{{6 l^2{\omega ^2}}} \nonumber\\
&+& \frac{{{2^{{1 \mathord{\left/
 {\vphantom {1 3}} \right.
 \kern-\nulldelimiterspace} 3}}}{c^2}{e^{ - i{\pi  \mathord{\left/
 {\vphantom {\pi  3}} \right.
 \kern-\nulldelimiterspace} 3}}}}}{{6{l^2}{\omega ^2}}}\frac{{\left( {1 - 6{l^2}{{\bf v}^2}{{\sin }^2}\theta } \right)}}{{{{\left[ {2 - 18{l^2}{{\bf v}^2}{{\sin }^2}\theta  - {l^2}\sqrt {324{{\bf v}^4}{{\sin }^4}\theta  + 432\frac{{{\omega ^2}}}{{{c^2}}}\left( {\frac{{{\omega ^2}}}{{{c^2}}} - {{\bf v}^2}} \right)} } \right]}^{{1 \mathord{\left/
 {\vphantom {1 3}} \right.
 \kern-\nulldelimiterspace} 3}}}}} \nonumber\\
 &+& \frac{{{c^2}{e^{i{\pi  \mathord{\left/
 {\vphantom {\pi  3}} \right.
 \kern-\nulldelimiterspace} 3}}}}}{{{2^{{1 \mathord{\left/
 {\vphantom {1 3}} \right.
 \kern-\nulldelimiterspace} 3}}}6{l^2}{\omega ^2}}}{\left[ {2 - 18{l^2}{{\bf v}^2}{{\sin }^2}\theta  - {l^2}\sqrt {324{{\bf v}^4}{{\sin }^4}\theta  + 432\frac{{{\omega ^2}}}{{{c^2}}}\left( {\frac{{{\omega ^2}}}{{{c^2}}} - {{\bf v}^2}} \right)} } \right]^{{1 \mathord{\left/
 {\vphantom {1 3}} \right.
 \kern-\nulldelimiterspace} 3}}}, \nonumber\\
\label{PodCFJ70}
\end{eqnarray}
where $\theta$ is the angle between $\bf v$ and $\bf k$.
Once again, the preceding electromagnetic vacuum behaves as a birefringent, dichroic, and dispersive medium.

Finally, obtaining the refractive indices for the CFJ model is straightforward. Again, if $\theta$ represents the angle between $\bf k$ and $\bf v$, we can easily obtain the refractive indices from equation (\ref{PodCFJ40}), that is,
\begin{equation}
{\left( {{\omega ^2} - {{\bf k}^2}} \right)^2} - {{\bf v}^2}\left( {{\omega ^2} - {{\bf k}^2}} \right) - {{\bf v}^2}{{\bf k}^2}{\cos ^2}\theta  = 0. \label{refr20}
\end{equation}
From this equation, it follows that 
\begin{equation}
{n^2}\left( \omega  \right) = 1 - \frac{{{{\bf v}^2}}}{{2{\omega ^2}}}{\sin ^2}\theta  \pm \frac{{|{\bf v}|}}{2}\sqrt {\frac{4}{{{\omega ^2}}} - \left( {\frac{2}{{{\omega ^2}}} - \frac{{{{\bf v}^2}}}{{{\omega ^4}}}{{\sin }^2}\theta } \right){{\sin }^2}\theta }.  \label{refr25}
\end{equation}

Before concluding this Section, one remark is pertinent at this point. This observation will help enhance our understanding of the formal structure of the theory we are studying.
First, it is worth noting that
the term involving high derivatives in Equation (\ref{PodCFJ01}) can be removed by introducing an auxiliary field, $Z^{\mu}$, according to the Lagrangian that follows:
\begin{equation}
{\cal L} =  - \frac{1}{4}F_{\mu \nu }^2 - \frac{1}{2}{Z^\mu }{Z_\mu } + l{Z^\mu }{\partial ^\alpha }{F_{\alpha \mu }} + \frac{1}{4}{\varepsilon ^{\mu \nu \kappa \lambda }}{v_\mu }{A_\nu }{F_{\kappa \lambda }}. \label{refr30}
\end{equation}
A partial integration in the mixed $Z^{\mu}-F_{\alpha\mu}$ term yields:
\begin{equation}
{\cal L} =  - \frac{1}{4}F_{\mu \nu }^2 - \frac{1}{2}{Z^\mu }{Z_\mu } - \frac{l}{2}{Z^{\alpha \mu }}{F_{\alpha \mu }} + \frac{1}{4}{\varepsilon ^{\mu \nu \kappa \lambda }}{v_\mu }{A_\nu }{F_{\kappa \lambda }}, \label{refr35}
\end{equation}
where ${Z_{\alpha \mu }} \equiv {\partial _\alpha }{Z_\mu } - {\partial _\mu }{Z_\alpha }$.
This Lagrangian can be brought into the form
\begin{equation}
{\cal L} =  - \frac{1}{4}{\left( {{F_{\mu \nu }} + l{Z_{\mu \nu }}} \right)^2} + \frac{{{l^2}}}{4}Z_{\mu \nu }^2 - \frac{1}{2}Z_\mu ^2 + \frac{1}{4}{\varepsilon ^{\mu \nu \kappa \lambda }}{v_\mu }{F_{\kappa \lambda }}, \label{refr40}
\end{equation}
which allows us to redefine new fields as below suitably:
\begin{equation}
F_{\mu \nu }^ \prime  \equiv {F_{\mu \nu }} + l{Z_{\mu \nu }}, \label{refr45-a}
\end{equation}
\begin{equation}
A_\mu ^ \prime  \equiv {A_\mu } + l{Z_\mu }, \label{refr45-b}
\end{equation}
\begin{equation}
Z_\mu ^ *  \equiv l{Z_\mu }. \label{refr45-c}
\end{equation}
In passing, we also note that $Z_{\mu}$ has mass dimension $2$, whereas $Z_{\mu}^{\prime}$ exhibits the usual canonical dimension of mass, like the photon field.

In terms of the primed fields defined above, the Lagrangian becomes
\begin{equation}
{\cal L} =  - \frac{1}{4}F_{\mu \nu }^{ \prime 2} + \frac{1}{4}Z_{\mu \nu }^{ \prime 2} - \frac{1}{{2{l^2}}}Z_\mu ^{ \prime 2} + \frac{1}{4}{\varepsilon ^{\mu \nu \kappa \lambda }}{v_\mu }A_\nu ^ \prime F_{\kappa \lambda }^ \prime  + \frac{1}{4}{\varepsilon ^{\mu \nu \kappa \lambda }}{v_\mu }Z_\mu ^ \prime Z_{\kappa \lambda }^ \prime  - \frac{1}{2}{\varepsilon ^{\mu \nu \kappa \lambda }}{v_\mu }A_\nu ^ \prime Z_{\kappa \lambda }^ \prime.  \label{refr50}
\end{equation}

If the LSV parameter were absent, the two vector modes, $A_{\mu}^{\prime}$ and  $Z_{\mu}^{\prime}$, would not mix. We could think of diagonalizing the symmetric matrix responsible for the mixing through an $SO(2)$-matrix. However, the kinetic terms and the $Z^{\prime}$-field mass term would yield new mixings between $A^{\prime}$ and $Z^{\prime}$. So, it is better to keep the kinetic and mass terms in diagonal form. We can conclude that the contemporary presence of the Podolsky and CFJ terms induces an oscillation mechanism between the photon, interpreted as the quantum associated with the field $A^{\prime}$, and the quantum of the  $Z^{\prime}$-field brought about by the higher-derivative term.

\section{Electromagnetic radiation}

We will now discuss the issue of obtaining the electromagnetic radiation emitted by a moving charged particle interacting with a medium characterized by a structure originating from Podolsky electromagnetism. 

Accordingly, the new Maxwell equations for a moving charged particle in a medium characterized by Podolsky electrodynamics are as follows:
\begin{eqnarray}
\left( {1 + {l^2}\, \Box } \right)\nabla  \cdot {\bf E} = \frac{{4\pi }}{\varepsilon }{\rho _{ext}}, \nonumber\\
\nabla  \cdot {\bf B} = 0, \nonumber\\
\nabla  \times {\bf E} + \frac{1}{c}\frac{{\partial {\bf B}}}{{\partial t}} = 0, \nonumber\\
\left( {1 + {l^2}\,\Box } \right)\left[ {\nabla  \times {\bf B} - \frac{{\varepsilon \mu }}{c}\frac{{\partial {\bf E}}}{{\partial t}}} \right] = \frac{{4\pi \mu }}{c}{J_{ext}}.  \label{Pod05}
\end{eqnarray}
Here, ${\rho _{ext}}$ and ${{\bf J}_{ext}}$ denote the external charge and current densities. 

It is worth noting that the equations above can also be expressed in the following form
\begin{widetext}
\begin{subequations}
\begin{eqnarray}
\left( {{\nabla ^2} - \frac{1}{{{c^{\prime 2}}}}\frac{{{\partial ^2}}}{{\partial {t^2}}}} \right){\bf E} = \frac{{4\pi }}{\varepsilon }\frac{1}{{\left( {1 + {l^2}\,\Box } \right)}}\nabla {\rho _{ext}} + \frac{{4\pi \mu }}{{{c^2}}}\frac{1}{{\left( {1 + {l^2}\,\Box } \right)}}\frac{{\partial {{\bf J}_{ext}}}}{{\partial t}}, \label{Pod10a}\\
\left( {{\nabla ^2} - \frac{1}{{{c^{\prime 2}}}}\frac{{{\partial ^2}}}{{\partial {t^2}}}} \right){\bf B} =  - \frac{{4\pi \mu }}{c}\frac{1}{{\left( {1 + {l^2}\,\Box } \right)}}\nabla  \times {{\bf J}_{ext}}, \label{Pod10b}
\end{eqnarray}
 \end{subequations}
 \end{widetext}
where the external charge and current densities are given by: ${\rho _{ext}}\left( {t,{\bf x}} \right) = Q\delta \left( x \right)\delta \left( y \right)\delta \left( {z - vt} \right)$ and ${\bf J}_{ext}\left( {t,{\bf x}} \right) = Qv\delta \left( x \right)\delta \left( y \right)\delta \left( {z - vt} \right){\hat {\bf e}_z}$. Here, we have simplified our notation by setting $\frac{1}{{c}^{\prime{2}}}\equiv\frac{\varepsilon\mathit{\mu}}{{c}^{2}}$.

Next, as we have indicated in \cite{Cherenkov23}, we Fourier transform to momentum space via
\begin{equation}
F(t,{\bf x}) = \int {\frac{{d \omega {d^3}{\bf k}}}{{{{\left( {2\pi } \right)}^4}}}} {e^{ - i\,\omega t + i\,{\bf k} \cdot {\bf x}}}F\left( {\omega,{\bf k}} \right), \label{Pod15}
\end{equation}
where $F$ stands for the electric and magnetic fields. 

Once this is done, we can derive the following expressions for the magnetic and electric fields: 
\begin{equation}
{\bf B}\left( {\omega ,{\bf k}} \right) = i\frac{{4\pi \mu }}{c}\frac{1}{{{\cal P}\,{\cal O}}}\left( {{\bf k} \times {{\bf J}_{ext}}} \right),  \label{Pod20}
\end{equation}
and
\begin{equation}
{\bf E}\left( {\omega ,{\bf k}} \right) =  - i\frac{{4\pi }}{\varepsilon}\frac{1} {{{\cal P}\,{\cal O}}}\,{\bf k}\,{\rho _{ext}} + i\frac{{4\pi \mu }}{{{c^2}}}\frac{1}{{{\cal P}\,{\cal O}}}\,\omega\, {\bf J}_{ext},  \label{Pod25}
\end{equation}
here $\cal P$ and $\cal O$ are expressed as  
\begin{equation}
{\cal P} = {{\bf k}^2} - \frac{{{\omega ^2}}}{{{c^{\prime 2}}}},  \label{Pod30}
\end{equation}
and
\begin{equation}
{\cal O} = 1 + {l^2}\left( {{{\bf k}^2} - \frac{{{\omega ^2}}}{{{c^2}}}} \right).  \label{Pod35}
\end{equation}

Similarly, the external charge and current densities in the Fourier space take the form: $\rho_{ext} \left( {w,{\bf k}} \right) = 2\pi Q\delta \left( {w - {k_z}v} \right)$ and ${{\bf J}_{ext}}\left( {w,{\bf k}} \right) = 2\pi Qv\delta \left( {w - {k_z}v} \right){\hat {\bf e}_z}$.

By proceeding in the same way as in \cite{Cherenkov23}, we shall now compute ${\bf B}\left( {w,{\bf x}} \right)$ and ${\bf E}\left( {w,{\bf x}} \right)$. In such a case, ${\bf B}\left( {w,{\bf x}} \right)$ is given by
\begin{equation} 
{\bf B}\left( {w,{\bf x}} \right) = \int {\frac{{{d^3}{\bf k}}}{{{{\left( {2\pi } \right)}^3}}}} \; {e^{i{\bf k} \cdot {\bf x}}}\;{\bf B}\left( {w,{\bf k}} \right). \label{Pod40}
\end{equation}

Making use of the problem's axial symmetry and using cylindrical coordinates, the magnetic field (\ref{Pod40}) reads
\begin{equation}
{\bf B}\left( {\omega ,{\bf x}} \right) =  - \frac{{iQ}}{{\pi c}}{e^{i{{\omega z} \mathord{\left/
 {\vphantom {{\omega z} v}} \right.
 \kern-\nulldelimiterspace} v}}}\int_0^\infty  {d{k_T}} {k_T}\int_0^{2\pi } {d\alpha } \frac{{{e^{i{k_T}{x_T}\cos \alpha }}}}{{{{\left. {\left( {{{\bf k}^2} - \frac{{{n^2}{\omega ^2}}}{{{c^2}}}} \right)} \right|}_{{k_z} = {\omega  \mathord{\left/
 {\vphantom {\omega  v}} \right.
 \kern-\nulldelimiterspace} v}}}}}\left[ {{k_T}sen\alpha\, \pmb {\hat \rho}  - {k_T}\cos \alpha\, \pmb {\hat \phi} } \right], \label{Pod45}
\end{equation}
where $\pmb{\hat \rho}$ and $\pmb{\hat \phi}$ are unit vectors normal and tangential to the cylindrical surface, respectively. 
 Here, the subscript $T$ in $k_{T}$ indicates transversal to the $z$ direction. We also notice that up to order $l^{2}$ we have used that $\frac{\mu }{{{\cal P}{\cal O}}} = \frac{1}{{\left( {{{\bf k}^2} - \frac{{{n^2}{\omega ^2}}}{{{c^2}}}} \right)}}$.

In addition, it can be noted that $\int_0^{2\pi } {d\theta } {e^{ix\cos \theta }}\sin \theta  = 0$, $\int_0^{2\pi } {d\theta } {e^{ix\cos \theta }}\cos \theta  = 2\pi i{J_1}\left( x \right)$,  where ${{J_1}\left( x \right)}$ is a  Bessel function of the first kind. With the aid of the previous integrals, we find that  equation (\ref{Pod45}) becomes
\begin{equation}
{\bf B}\left( {\omega ,{\bf x}} \right) =  - \frac{{2Q}}{c}{e^{i{{\omega z} \mathord{\left/
 {\vphantom {{\omega z} v}} \right.
 \kern-\nulldelimiterspace} v}}}\int_0^\infty  {d{k_T}k_T^2} \frac{{{J_1}\left( {{k_T}{x_T}} \right)}}{{\left( {{\bf k}_T^2 - {\alpha ^2}} \right)}}\,\pmb{\hat \phi}, \label{Pod50}  
\end{equation}
where ${\alpha ^2} = \frac{{{\omega ^2}{n^2}}}{{{c^2}}}\left( {1 - \frac{{{c^2}}}{{{n^2}{v^2}}}} \right)$.

Then, by integrating over $k_T$ and performing further manipulations, we find that equation (\ref{Pod50}) may be rewritten as
\begin{equation}
{\bf B}\left( {\omega ,{\bf x}} \right) = \frac{{i\pi Q}}{c}{e^{i{{\omega z} \mathord{\left/
 {\vphantom {{\omega z} v}} \right.
 \kern-\nulldelimiterspace} v}}}\alpha H_1^{\left( 1 \right)}\left( {\alpha {x_T}} \right)\pmb{\hat \phi}, \label{Pod55}
\end{equation}
here we have used ${x_T} = \rho$ (cylindrical coordinates). We also have that
${H}_{1}^{(1)}(x)$ is a Hankel function of the first kind.

Our next task is to calculate the electric field. To do this, we begin noting that up to order  $l^{2}$ we can express $\frac{1}{{\varepsilon {\cal P}{\cal O}}}$ as $\frac{1}{{\varepsilon {\cal P}{\cal O}}} = \frac{1}{{\left( {{{\bf k}^2} - \frac{{{n^2}{\omega ^2}}}{{{c^2}}}} \right)}} - 2{l^2}\frac{{\left( {{{\bf k}^2} - \frac{{{\omega ^2}}}{{{c^2}}}} \right)}}{{\left( {{{\bf k}^2} - \frac{{{n^2}{\omega ^2}}}{{{c^2}}}} \right)}}$. 
In such a case, by applying the same method as previously, we can derive from the expression (\ref{Pod25}) that the electric field becomes\\
\begin{eqnarray}
{\bf E}\left( {\omega ,{\bf x}} \right) &=&  - \frac{{iQ}}{{\pi v}}{e^{i{{\omega z} \mathord{\left/
 {\vphantom {{\omega z} v}} \right.
 \kern-\nulldelimiterspace} v}}}\left\{ {2\pi i\int_0^\infty  {d{k_T}k_T^2\frac{{{J_1}\left( {{k_T}{x_T}} \right)}}{{\left( {k_T^2 - {\alpha ^2}} \right)}} \pmb {\hat \rho}  + \frac{{2\pi \omega }}{v}\int_0^\infty  {d{k_T}} {k_T}\frac{{{J_0}\left( {{k_T}{x_T}} \right)}}{{\left( {k_T^2 - {\alpha ^2}} \right)}} \pmb{{{\hat e}_z}} }} \right\} \nonumber\\
 &+& \frac{{i2{l^2}Q}}{{\pi v}}{e^{i{{\omega z} \mathord{\left/
 {\vphantom {{\omega z} v}} \right.
 \kern-\nulldelimiterspace} v}}}\left\{ {2\pi i\int_0^\infty  {d{k_T}k_T^2} \frac{{\left( {k_T^2 + \frac{{{\omega ^2}}}{{{v^2}}} - \frac{{{\omega ^2}}}{{{c^2}}}} \right)}}{{\left( {k_T^2 - {\alpha ^2}} \right)}}{J_1}\left( {{k_T}{x_T}} \right) \pmb {\hat \rho} } \right\} \nonumber\\
&+& \frac{{i2{l^2}Q}}{{\pi v}}{e^{i{{\omega z} \mathord{\left/
 {\vphantom {{\omega z} v}} \right.
 \kern-\nulldelimiterspace} v}}}\left\{ {\frac{{2\pi \omega }}{v}\int_0^\infty  {d{k_T}{k_T}\frac{{\left( {k_T^2 + \frac{{{\omega ^2}}}{{{v^2}}} - \frac{{{\omega ^2}}}{{{c^2}}}} \right)}}{{\left( {k_T^2 - {\alpha ^2}} \right)}}{J_0}\left( {{k_T}{x_T}} \right) \pmb{{{\hat e}_z}}}} \right\} \nonumber\\
&+&\frac{{i2\omega Q}}{{{c^2}}}{e^{i{{\omega z} \mathord{\left/
 {\vphantom {{\omega z} v}} \right.
 \kern-\nulldelimiterspace} v}}}\int_0^\infty  {d{k_T}{k_T}\frac{{{J_0}\left( {{k_T}{x_T}} \right)}}{{\left( {k_T^2 - {\alpha ^2}} \right)}}}  \pmb{{{\hat e}_z}}, \label{Pod60}                                  
\end{eqnarray}
where $\pmb {{{\hat e}_z}}$ is a unit vector along the $z$ direction
and we utilized the equation $\int_0^{2\pi } {d\theta } {e^{ix\cos \theta }} = 2\pi {J_0}\left( x \right)$, where ${{J_0}\left( x \right)}$ is a Bessel function of the first kind. Again, by integrating over $k_T$, equation (\ref{Pod60}) can be expressed as: 
\begin{eqnarray}
{\bf E}\left( {\omega ,{\bf x}} \right) &=&  - i\pi Q{e^{i{{\omega z} \mathord{\left/
 {\vphantom {{\omega z} v}} \right.
 \kern-\nulldelimiterspace} v}}}\frac{\alpha }{v}H_1^{\left( 1 \right)}\left( {\alpha {x_T}} \right) \pmb {\hat \rho}  - i2\pi {l^2}Q{e^{i{{\omega z} \mathord{\left/
 {\vphantom {{\omega z} v}} \right.
 \kern-\nulldelimiterspace} v}}}\frac{\alpha }{v}\left[ {\frac{{{\omega ^2}\left( {{n^2} - 1} \right)}}{{{c^2}}} - \frac{{2{\omega ^2}}}{{{v^2}}}} \right]H_1^{\left( 1 \right)}\left( {\alpha {x_T}} \right)\pmb  {\hat \rho} \nonumber\\
 &+& \pi Q{e^{i{{\omega z} \mathord{\left/
 {\vphantom {{\omega z} v}} \right.
 \kern-\nulldelimiterspace} v}}}\frac{{{\alpha ^2}}}{\omega }H_0^{\left( 1 \right)}\left( {\alpha {x_T}} \right)\pmb {{{\hat e}_z}} + \frac{{\pi Q}}{{{l^2}\omega }}{e^{i{{\omega z} \mathord{\left/
 {\vphantom {{\omega z} v}} \right.
 \kern-\nulldelimiterspace} v}}}H_0^{\left( 1 \right)}\left( {\alpha {x_T}} \right) \pmb{{{\hat e}_z}}   \nonumber\\
 &-& 4\pi {l^2}Q{e^{i{{\omega z} \mathord{\left/
 {\vphantom {{\omega z} v}} \right.
 \kern-\nulldelimiterspace} v}}}\left[ {{\alpha ^2} + \frac{1}{2}\left( {\frac{{{\omega ^2}}}{{{c^2}}} - \frac{{{\omega ^2}}}{{{v^2}}}} \right)} \right]H_0^{\left( 1 \right)} \pmb{{{\hat e}_z}}.  \label{Pod65}
\end{eqnarray}

With the information provided, we can now calculate the radiated energy in the current scenario. To do so, we will estimate the radiated energy by calculating the Poynting vector. In particular, we will focus on the power density carried by the radiation fields across the surface denoted as $S$, which bounds the volume $V$. This power density is represented by the real part of the time-averaged Poynting vector,
\begin{equation}
{\bf S} = \frac{c}{{2\pi }}{\mathop{\rm Re}\nolimits} \left( {{\bf E} \times {{\bf B}^ * }} \right). \label{Pod70}
\end{equation}

To be more precise, we will calculate the power radiated through the surface $S$ \cite{Das}, as follows
 \begin{equation}
{\cal E} = \int_{ - \infty }^\infty \! {dt} \int\limits_S {d{\bf a} \cdot {\bf S}}.  \label{Pod75}
\end{equation}

With this in mind, let us consider a cylinder with length $h$ and radius $\rho$.
Therefore, the total energy radiated through the surface of the cylinder 
can be expressed as 
\begin{equation}
{\mathcal{E}} = \frac{c}{{2\pi }}\left( {2\pi \rho h} \right)\int_0^\infty  {d\omega }\, {S_\rho }\,\Theta \left( {v - \frac{c}{n}} \right). \label{Pod80}
\end{equation}
It is worth noting that, from our previous discussion \cite{Cherenkov23}, we included the step function to account for Cherenkov radiation being emitted only if the velocity of the particle exceeds the velocity of light in the medium (${v} > \frac{c}{n}$).

Using equations (\ref{Pod55}) and (\ref{Pod65}) in the radiation zone, we can express the power radiated per unit length (\ref{Pod80}) as
\begin{equation}
W \equiv \frac{\cal E}{h} = - \frac{{2\pi {Q^2}}}{{{c^2}}}\int_0^\infty  {d\omega }\, {n^2}\left( \omega  \right)\omega \left( {1 - \frac{{{c^2}}}{{{n^2}{v^2}}}} \right){e^{ - 2 \rho\,\frac{{\omega n\left( \omega  \right)}}{c} \sqrt {\frac{{{c^2}}}{{{n^2}\left( \omega  \right){v^2}}} - 1} }}. \label{Pod85}
\end{equation}
It is important to note that the expression above is similar to the one obtained in the theory of Cherenkov radiation, except for the additional factors $n^2(w)$ and ${e^{ - 2\rho\,\frac{{\omega n\left( \omega  \right)}}{c}\sqrt {\frac{{{c^2}}}{{{n^2}\left( \omega  \right){v^2}}} - 1} }}$ in the integrand of equation (\ref{Pod85}).

\section{Concluding Remarks}

In our concluding remarks, we would like to provide additional justification for our investigation into the presence of the higher-derivative Podolsky term and the LSV contribution described by the Carroll-Field-Jackiw (CFJ) four-vector, $v_\mu$. Both terms can be viewed as being generated as effective contributions through one-loop integration over fermions coupled to the photon field. One important effect we should point out is the possibility that the CFJ action term induces a mixing between the photon and the massive vector mode introduced by the Podolsky term. In the present paper, we have only checked that this photon-massive vector takes place. We are next going to inspect this phenomenon within the context of the Euler-Heisenberg non-linear electrodynamic model, having in mind that integration upon the electron-positron field induces, at the 1-loop level, the simultaneous presence of Podolsky, CFJ, and higher-power terms of the electromagnetic field. As a future prospect, we intend to endeavor a complete study of electromagnetic radiation in this more complete scenario, especially because of the photon-massive vector mixing. 

On the other hand, if we look back at eq. (\ref{refr50}), if we choose to diagonalize the photon-$Z ^{\prime}$ mixed CFJ action, the kinetic photon and $Z ^{\prime}$ terms get mixed. This may signal the direction of considering the massive model carried by the Podolsky term as a paraphoton and, consequently, if dark-sector matter is coupled to the system, this may induce the coupling of the photon to the dark-sector matter; this coupling is known as milli-charge. Pursuing an investigation of electromagnetic radiation in this scenario may unveil interesting features of the radiation produced by milli-charged fermions.\\

{\bf Acknowledgments}: 

One of us (P. G.) acknowledges the financial support received by ANID PIA/APOYO AFB230003. J. P. F. expresses deep gratitude to UNIFEI for the invaluable grant, to USM for their warm hospitality, and to CAPES for the financial support.


\begin{thebibliography}{99}
\bibitem{Bopp} F. Bopp, Ann. Phys. (Paris) {\bf 38}, 345 (1940).

\bibitem{Podolsky} B. Podolsky, Phys. Rev. {\bf 62}, 68 (1942).

\bibitem{Schwed} B. Podolsky and P. Schwed, Rev. Mod. Phys. {\bf 20}, 4 (1948).

\bibitem{Galvao} C. A. P. Galv\~ao and B. M. Pimentel, Can. J. Phys. {\bf 66}, 460 (1988).

\bibitem{Gaete} P. Gaete, Int. J. Mod. Phys. A {\bf 27}, 1250061, (2012).

\bibitem{Accioly} A. Accioly, P. Gaete, J. Helay\"el-Neto, E. Scatena and R. Turcati, Mod. Phys. Lett. A {\bf 2} 6, 1985 (2011).

\bibitem{West} P. West, Nucl. Phys. B {\bf 268}, 113 (1986).

\bibitem{Polyakov} A. Polyakov, Nucl. Phys. B {\bf 268}, 406 (1986).

\bibitem{Lee} T. Lee and G. Wick, Nucl. Phys. B {\bf 9}, 209 (1969).

\bibitem{Wick} T. Lee and G. Wick, Phys. Rev. D {\bf 2}, 1033 (1970).

\bibitem{Grinstein} B. Grinstein, D. O’Connell, and M. Wise, Phys. Rev. D {\bf 77}, 025012 (2008).

\bibitem{GaeteSpa} P. Gaete, and E. Spallucci, J. Phys. A {\bf 45}, 065401 (2012).

\bibitem{Donoghue} J. F. Donoghue, E. Golowich, and B. R. Holstein \textit{ Dynamics of the Standard Model}, (Cambridge University Press, 1992).

\bibitem{Euler} H. Euler and W. Heisenberg, Z. Phys. {\bf 98}, 714 (1936).

\bibitem{Adler} S. L. Adler, Ann. Phys. (N.Y.) {\bf 67}, 599 (1971).

\bibitem{Costantini} V. Costantini, B. De Tollis and G. Pistoni, Nuovo Cimento A {\bf 2}, 733 (1971).

\bibitem{Biswas} S. Biswas and K. Melnikov, Phys. Rev. D {\bf 75}, 053003 (2007).

\bibitem{Tommasini} D. Tommasini, A. Ferrando, H. Michinel and M. Seco, J. High Energy Phys. {\bf 0911}, 043 (2009).

\bibitem{Ferrando} A. Ferrando, H. Michinel, M. Seco and D. Tommasini, Phys. Rev. Lett. {\bf 99}, 150404 (2007).

\bibitem{Bamber} C. Bamber et al., Phys. Rev. D {\bf 60}, 092004 (1999).

\bibitem{Burke} D. L. Burke et al., Phys. Rev. Lett. {\bf 79}, 1626 (1997).

\bibitem{Pike} O. J. Pike, F. Mackenroth, E. G. Hill and S. J. Rose, Nature Photonics {\bf 8}, 434 (2014). 

\bibitem{Ejlli} A. Ejlli, F. Della Valle, U. Gastaldi, G. Messineo, R. Pengo, G. Ruoso, and G. Zavattini, Phys. Rept. {\bf 871}, 1--74 (2020).

\bibitem{Aaboud} M. Aaboud et al. (ATLAS Collaboration), Evidence for light-by-light scattering in heavy-ion collisions with the ATLAS detector at the LHC, arXiv:1702.01625. Published in {\bf Nature Physics} (2017).

\bibitem{Enterria} D. d'Enterria and G. G. da Silveira, Phys. Rev. Lett. {\bf 111}, 080405 (2013); Erratum, Phys. Rev. Lett. {\bf 116}, 129901(E) (2016).

\bibitem{Kostelecky1} V. A. Kostelecky, Phys. Rev. D {\bf 69}, 105009 (2004). 

\bibitem{Kostelecky2} V. A. Kostelecky and N. Russell, Rev. Mod. Phys. {\bf 83}, 11 (2011).

\bibitem{Samuel1} V. A. Kostelecky and S. Samuel, Phys. Rev. Lett. {\bf 63}, 224 (1989).

\bibitem{Samuel2} V. A. Kostelecky and S. Samuel,  Nucl. Phys. B {\bf 336}, 263 (1990).

\bibitem{Gambini} R. Gambini and J. Pullin, Phys. Rev. D {\bf 59}, 124021 (1999).

\bibitem{Alfaro} J. Alfaro, H. A. Morales-Tecotl and L. F. Urrutia,  Phys. Rev. D {\bf 65}, 103509 (2002).

\bibitem{Tony} A. Accioly and E. Scatena, Mod. Phys. Lett. A {\bf 25} 4, 269 (2010).

\bibitem{Cherenkov23}  P. Gaete and J. Helay\"el-Neto, Eur. Phys. J. C {\bf 84} 6, 609 (2024).

\bibitem{Das} A. Das, \textit{ Lectures on Electromagnetism}, (Hindustan Book Agency, 2004).
\end{thebibliography}
\end{document}